\documentclass[12pt]{emulateapj}

\usepackage[dvips]{color}

\shorttitle{Plasma composition in Cygnus A jet}
\shortauthors{Kino, Kawakatu, \& Takahara}
\begin{document}
\title{Calorimetry of Active Galactic Nucleus  jets:
testing plasma composition in Cygnus~A}
\author{M.Kino\altaffilmark{1},
  N.Kawakatu\altaffilmark{2},
  and 
  F.Takahara\altaffilmark{3}}
\altaffiltext{1}{National Astronomical Observatory of Japan
                 2-21-1 Osawa, Mitaka, Tokyo, 181-8588, Japan}
\altaffiltext{2}{
Graduate School of Pure and Applied Sciences,
University of Tsukuba,  305-8571 Tsukuba, Japan}
\altaffiltext{3}{Department of Earth and Space Science, 
Osaka University, 560-0043 Toyonaka, Japan}

\begin{abstract}

We examine plasma composition of jets in active galactic nuclei
through the comparison of the total pressure ($P$)
with partial pressures of electrons and protons in a cocoon.
The total pressure is estimated from the analysis of 
an expanding cocoon dynamics.
We  determine the average kinetic energy
per particle 
for several representative cases of particle energy
distribution such as one- and two-temperature
thermal plasmas and non-thermal electrons
by evaluating the
dissipation of total kinetic energy of the jet 
into the internal energy of cocoon plasma. 
The number density of the total 
electrons/positrons ($n_{\pm}$) in the cocoon
is constrained by using 
the particle supply from hot spots 
and the absence of thermal bremsstrahlung emission 
from radio lobes.
By inserting $P$, $n_{\pm}$ and 
the particle energy of each population
into the equation of state, the number density ($n_{p}$) 
and pressure ($P_{p}$) of protons in the cocoon can be constrained.
Applying this method to Cygnus A, 
we find that 
(i) electron/positron ($e^{\pm}$) pairs  
always dominate in terms of number density,
but that 
(ii) either an ``$e^{\pm}$-supported cocoon (i.e., $P_{\pm} >P_{p}$)'' 
or ``proton-supported one (i.e, $P_{\pm} <P_{p}$)'' 
is possible.

\end{abstract}

\section{INTRODUCTION}

Elucidating the formation mechanism 
of relativistic jets in active galactic nuclei (AGNs)
is one of the greatest challenges of astrophysics
in this century 
(e.g., 
Blandford and Znajek 1977;
McKinney 2006; 
Komissarov et al. 2007; 
McKinney et al. 2012). 
Plasma composition of jets 
is a  fundamental but difficult issue
(Begelman et al. 1984 for review),
because emission timescales 
of the bulk population such as 
low-energy electrons/positrons and 
protons are too long.
To examine plasma composition,
discrete blobs in blazar jets have
been utilized over the years.
So far, three approaches have been
pursued. 
The first is based on the synchrotron self-absorption limit
combined with total kinetic powers of jets  
(Reynolds et al. 1996; Hirotani et al. 1999,  2000; Hirotani 2005).
The literature indicates 
the existence of  $e^{\pm}$ pair plasma in M 87, 3C 279 and 3C 345.
The second is the constraint by the detection of
circular polarization.
Wardle et al. (1998) and Homan et al. (2009)
examined the case of 3C 279 and
found that the minimum Lorentz factor
of non-thermal electrons/positrons should be
much larger than unity for electron-proton
(hereafter $e/p$) content. 
They rather favored an alternative possibility of 
dominant  $e^{\pm}$ pair content
with a small minimum Lorentz factor of non-thermal 
electrons/positrons 
(see, however, Ruszkowski and Begelman 2002). 
The third approach is the constraint from the absence of 
bulk-Compton emission in flat spectrum radio quasars 
(Sikora and Madejski 2000; Ghisellini \& Tavecchio 2010) and 
it has been observationally tested 
for PKS~1510-089 and SWIFT~J0746.3+2548 
(Kataoka et al. 2008; Watanabe et al. 2009).
The same approach has also been applied  
to the kiloparsec-scale knots in PKS~~0637-752
(Georganopoulos et al. 2005; Uchiyama et al. 2005; Mehta et al. 2009).
They claim that jets contain more $e^{+}e^{-}$ pairs than protons,
but that jets are dynamically  dominated by protons. 
However, it should be noted
that the estimate of a total kinetic power $L_{\rm j}$ 
of each blob is difficult, because of the existence of
invisible components 
such as low-energy electrons/positrons and protons. 
Therefore, the assumption of
constant $L_{\rm j}$ was made 
and the $L_{\rm j}$  
are inferred from non-thermal emissions.
Since plasma composition is sensitive to $L_{\rm j}$, 
a better estimate of $L_{\rm j}$ is essential.
Regarding to the estimate of $L_{\rm j}$, 
it is essential to take into account of the
thermal component (e.g., Kino and Takahara 2008).

Cocoons  associated with
Fanaroff-Riley~I and ~II (FR~I and FR~II) radio galaxies
are also known to be good tools for exploring  plasma
composition.
In contrast to blobs in blazars,
investigations using cocoon dynamics
allow us to better estimate of  
energy injection into the cocoon.
The total pressure $P$ can be estimated with fewer uncertainties
based on the dynamical interaction between
jets and the intra-cluster medium (ICM) 
and $P$ involves the contributions of invisible components
(e.g., Rawlings and Sanders 1991; Fabian et al. 2002).
For FR~I radio galaxies, many authors have discussed 
the ratio of $P$ to 
that of non-thermal electrons ($P_{-}^{\rm NT}$) 
for various sources based on observed non-thermal emissions
(e.g., Dunn et al. 2005; Croston et al. 2005;
Rafferty et al. 2006;  De Young 2006; B{\^i}rzan  et al. 2008).
First of all, we should emphasize
that these studies indicate that
the total pressure $P$ tends to be larger than 
that of non-thermal electrons, 
i.e.,  $P>P_{-}^{\rm NT}$  
(e.g., B{\^i}rzan  et al. 2008).
This means that the finite pressure of 
low-energy electrons/positrons and/or protons
is required in these sources.
The derived $P/P_{-}^{\rm NT}$ values in the previous work
extend over a wide range 
from the order of unity to thousands
(e.g., Birzan et al. 2008; Cavagnolo et al. 2010).
For FR~I sources, however, an entrainment process of 
surrounding medium via the jet boundary layer could work
 (e.g., De Young 1993; Bicknell 1984; Rossi et al. 2008) 
and the process  makes jets heavier.
Therefore, jets in FR~I sources could undergo severe proton loading 
during their propagations
and this could cause the large scatter of 
$P/P_{-}^{\rm NT}$.

\begin{figure} 
\includegraphics[width=10cm]{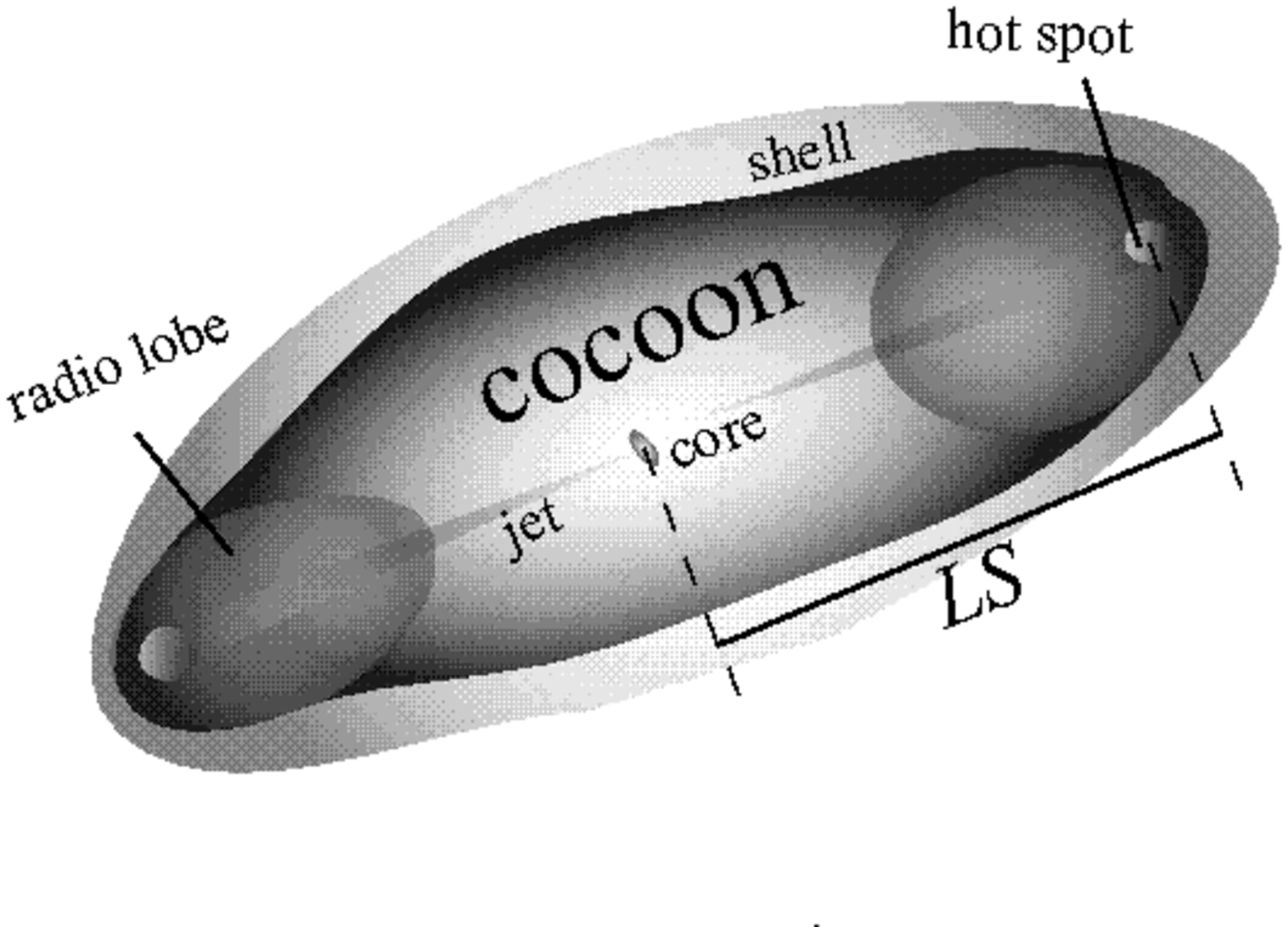}
\caption
{ A cartoon of a powerful FR II radio galaxy.
A pair of jets is ejected from the core and they
are decelerated via strong shocks. The shocks are identified as
the hot spots and the remnant of decelerated jets 
envelopes the overall jet system and this is identified as a cocoon.
Part of cocoon is normally observed as radio lobes.
The cocoon head and the hot spots advance at a speed $v_{\rm hs}$.
Swept-up ambient matter becomes 
a shell and surrounds the cocoon.
The projected linear size is denoted as $LS$ in this work.}
\label{fig:cocoon}
\end{figure}

Instead, in this work, we focus on 
FR~II radio galaxies (Fig. \ref{fig:cocoon})
from the viewpoint of the important advantage
they represent.
Contrary to FR~I sources,
we know from relativistic hydrodynamic simulations 
that no significant entrainment appears for 
FR~II sources (Scheck et al. 2002; Mizuta et al. 2004).
Therefore, a plasma composition test for 
FR~II radio galaxies  would allow us to give 
better constraints on plasma composition in AGN jets
without an entrainment effect. 
Regarding an observational indication 
of a difference between total and non-thermal pressures 
in FR~II radio galaxies,
Ito et al. (2008) (hereafter I08) recently
examined for the following sources
(Cygnus A, 3C~223, 3C~284 and 3C~219).
In I08, they show that the energy density of total plasma
is larger than the energy density of non-thermal electrons
by the factor of 4-310 in the case of minimum-energy condition 
(e.g., Miley 1980; Kellermann and Pauliny-Toth 1981).
This implies that the minimum-energy condition 
is violated, particle energy is dominant, 
and low-energy electrons/positrons and/or
protons (i.e., cosmic-rays) are required to  
explain the total $P$ in these FR~II sources.

In \S 2, we describe the basic idea and assumptions 
of our method.
In \S 3, we briefly explain the dynamical determination
of the total pressure in the cocoon.
In \S 4 we express $P$ as functions of 
the number density ratio of protons to electrons. 
In \S 5, we explain details of the plasma composition test. 
It is applied to Cygnus A in \S 6. 
Summary and discussions are given in \S 7.

\section{Method and problem setting}

Here, we describe the basic idea
and assumptions of our method.
In this work,
the number densities of 
protons ($n_{p}$), positrons ($n_{+}$) 
and electrons ($n_{-}$) are related using 
the parameter $\eta$ as follows:
\begin{eqnarray} \label{eq:eta}
n_{p}&\equiv&\eta n_{-}   \nonumber \\
n_{+}&=&      (1-\eta)n_{-}  \quad (0\leq \eta\leq 1)    ,
\end{eqnarray}
where the latter relation is derived from
the charge neutrality condition.
The case of $\eta=0$ corresponds to pure $e^{\pm}$ plasma
while  $\eta=1$ corresponds to the pure $e/p$ plasma.
We denote that  
$n_{p}=n_{p}^{\rm T}+n_{p}^{\rm NT}$, 
$n_{-}=n_{-}^{\rm T}+n_{-}^{\rm NT}$,  
$n_{+}=n_{+}^{\rm T}+n_{+}^{\rm NT}$, and
$n_{\pm}=n_{-}+n_{+}$
where $n_{\pm}$ is the sum of the total 
number densities of electrons and positrons. 
Hereafter, superscripts T and NT represent thermal
and non-thermal components, respectively.
The distinction between thermal and non-thermal particles
may not be trivial for relativistic plasmas.
In this paper, we refer the thermal component to
Maxwellian-like distribution characterized by the temperature,
while we refer the non-thermal one to particles 
following a power-law distribution characterized
by the power-law index and minimum and maximum energies
as detailed below.
Since we focus on relativistic plasmas in the present work,
thermal component correspondingly has a relativistic temperature.
Hence one should be cautious since most of 
observational papers refer the thermal component to 
non-relativistic plasmas (e.g., Garrington and Conway 1991).

The allocation of partial pressure of each 
plasma population is the central concern of this paper. 
In general, $P$ is decomposed to
\begin{eqnarray}\label{eq:p-def}
P&=&P_{-}+P_{+}+P_{p}+P_{\rm B} \nonumber \\ 
&=&
P_{-}^{\rm T}+P_{+}^{\rm T}
+ P^{\rm NT}_{-} + P^{\rm NT}_{+}+ P_{p}^{\rm T}+ P^{\rm NT}_{p} 
+P_{\rm B},
\end{eqnarray}
where 
$P^{\rm T}_{-}$,
$P^{\rm T}_{+}$,
$P^{\rm T}_{p}$,
$P^{\rm NT}_{-}$,
$P^{\rm NT}_{+}$,
$P^{\rm NT}_{p}$, and
$P_{\rm B}$ are, the partial pressures of 
thermal (T) electrons,
thermal positrons,
thermal protons,
non-thermal (NT) electrons,
non-thermal positrons,
non-thermal protons, and 
a magnetic pressure respectively.
We also define 
$P_{\pm}=P_{-}+P_{+}$ as the sum of the total 
pressures of electrons and positrons. 
Throughout this work,
we do not include the magnetic pressure 
because it is sub-dominant in $P$.
Isobe et al. (2005) summarize 
the energy density of energetic electrons as typically being
10 times larger than that of magnetic fields in various radio lobes
(e.g., 
Isobe et al. 2002; 
Tashiro et al. 1998, 2009;
Hardcastle and Croston 2010) and
it also holds in Cygnus A (Yaji et al. 2010).

\subsection{Basic idea of the method}

The essence of our method is as follows.
First,
the total pressure in the cocoon ($P$) 
is determined through dynamical considerations 
following I08 where they obtained $P$ via 
the comparison of the expanding cocoon model 
with radio observations.
Second, average energy per one particle
in the cocoon is evaluated.
It is essential that
our formulation is based on the basic conservation laws
of mass, momentum, and energy in the cocoon.
Since it depends on coupling of protons 
to the electrons/positrons,
we examine several  representative cases with 
different equations of state.
Third, $n_{-}$ can be partially constrained 
by using the absence of thermal bremsstrahlung emission
from the cocoon
and the supply rate of electrons from the hot spots.
Finally, $n_{p}$ and $P_{p}$ can be 
obtained by inserting the obtained quantities into 
the equation of state (EOS).

\subsection{On particle distribution functions}

Since observational data at
low frequencies below GHz are quite limited,
it is hard to explore the properties 
of low-energy electrons 
(including positrons).
Bearing this difficulty in mind,
we pick up plausible cases of 
electron distribution function.
As the canonical case referred to as case (a),
we consider two-temperature thermal  plasmas, 
where protons and electrons have different temperatures
and contributions of non-thermal components 
to the total pressure are negligible. 
As an alternative,
we also examine case (b) where protons and electrons
take the same temperature
without non-thermal components.

We further explore two cases (c) and (d)
in which 
non-thermal population makes a dominant contribution
to the total pressure 
with a negligible pressure of thermal population. 
For the non-thermal population, 
we assume the powe- law distribution functions:
\begin{eqnarray}\label{eq:ne-c}
n_{-}^{\rm NT}(\gamma_{-}) &\propto& \gamma_{-}^{-s_{e}}
(\gamma_{-,\rm min}\le \gamma_{-} \le \gamma_{-,\rm max}), \nonumber \\
n_{p}^{\rm NT}(\gamma_{p}) &\propto& \gamma_{p}^{-s_{p}}
(\gamma_{p,\rm min}\le \gamma_{p} \le \gamma_{p,\rm max}),
\end{eqnarray}
for case (c) with $s_{p}=s_{e}>2$.
Observations of the spectral index in the 
radio lobe of Cygnus A suggest $s_{e}>2$ 
(e.g., Carilli et al. 1991; Yaji et al. 2010).

Lastly, we set case (d) 
in which the number spectrum of non-thermal
electrons is given by a broken power law:
\begin{eqnarray}\label{eq:ne-d}
n_{-}^{\rm NT}(\gamma_{-}) &\propto&
\left\{ \begin{array}{ll}
\gamma_{-}^{-s_{e,1}}
& \mbox{$(\gamma_{-,\rm min}\le \gamma_{-} \le \gamma_{-,\rm crit})$}, \\
\gamma_{-,\rm crit}^{s_{e,2}-s_{1}}\gamma_{-}^{-s_{e,2}}
& \mbox{$(\gamma_{-,\rm crit}\le \gamma_{-} \le \gamma_{-,\rm max})$},
\end{array} \right.   \nonumber \\
n_{p}^{\rm NT}(\gamma_{p}) &\propto& \gamma_{p}^{-s_{p}}
(\gamma_{p,\rm min}\le \gamma_{p} \le \gamma_{p,\rm max}),
\end{eqnarray}
where $s_{e,1}<2$ and $s_{p}>2$ are satisfied. 
This model is based on Stawarz et al. (2007)
who suggest that observed spectra
at the jet termination shock (hot spot) of FR II jets (Cygnus A)
can be explained by the break at non-thermal electron energy 
(hereafter $\gamma_{\pm,\rm crit}$).
This type of spectra may be
due to the absorption of electromagnetic waves emitted 
at the harmonics of cyclotron frequency of cold protons,
as discussed by Hoshino et al. (1992)
and Amato and Arons (2006).
Some observations for other FR~II sources
could also be compatible with
this picture (e.g., Perlman et al. (2010) for 3C445.)

For cases (c) and (d),
the minimum energy of non-thermal electrons/positrons
($\gamma_{\pm,\rm min}m_{e}c^{2}$) and protons 
($\gamma_{p,\rm min}m_{e}c^{2}$) are 
generally assumed as
\begin{eqnarray} 
\gamma_{\pm,\rm min}\approx
\gamma_{p,\rm min}\approx \Gamma_{\rm j} ,
\end{eqnarray}
which is expected when protons and electrons/positrons 
are separately heated and accelerated 
at termination shocks.
On the other hand,
the values of
the maximum energy of non-thermal pairs
($\gamma_{\pm,\rm max}m_{e}c^{2}$)
and protons ($\gamma_{p,\rm max}m_{p}c^{2}$ )
are largely uncertain. 
While $\gamma_{\pm,\rm max}m_{e}c^{2}$ may be 
significantly affected by radiative coolings, 
$\gamma_{p,\rm max}m_{p}c^{2}$
may reach the range of highest energy cosmic-rays
(e.g., 
Takahara 1990; 
Rachen and Biermann 1993).
It is reasonable to suppose that 
$\gamma_{\pm,\rm max}\gg \gamma_{\pm,\rm min}$
and $\gamma_{p,\rm max}\gg \gamma_{p,\rm min}$.

\section{Total pressure $P$}

In this section, we briefly describe 
the basic idea of estimating the total pressure $P$.
In Fig. \ref{fig:cocoon} we show a cartoon of the interaction 
of the jet and ICM.
Heating and acceleration processes work at hot spots 
and those particles are injected into cocoons.
The cocoon model was proposed by Begelman and Cioffi (1989)
in which the dissipated energy of jet bulk motion is the origin of
the total pressure of cocoon 
and 
a cocoon of FR IIs is expected to be 
overpressured against ICM pressure ($P_{\rm ICM}$)
with a significant sideways expansion.
Therefore, the assumption of $P=P_{\rm ICM}$ is not valid.
We have proposed the method of dynamical constraint on $P$ 
by comparison of the cocoon model with the actually
observed morphology of the cocoons (Kino and Kawakatu 2005; I08)
and the method is applied to various radio lobes 
(e.g., Machalski et al. 2010).
We use this model in the present work.
The reliability of the expanding cocoon model is well examined 
in Kawakatu and Kino (2006).
The results of relativistic hydrodynamical simulations of
Scheck et al. (2002) and Perucho and Marti (2007)
support the above analytic model.
The 
mass and energy injections from the  jet into
the cocoon, which govern 
the cocoon pressure $P$ and 
mass density $\rho$ 
averaged by the source age ($t_{\rm age}$)
are written as
\begin{eqnarray}\label{eq:pc}
\frac{{\hat \gamma}}{{\hat \gamma}-1}
\frac{PV}{t_{\rm {age}}}=
2 T^{01}_{\rm j}  
A_{\rm j} \equiv 2 L_{\rm j}, \quad
T_{\rm j}^{01}=\rho_{\rm j}c^{2}\Gamma_{\rm j}^{2}v_{\rm j}  ,
\end{eqnarray}
\begin{eqnarray}\label{eq:rho}
\frac{\rho V}{t_{\rm age}}=
2 J_{\rm j}  
A_{\rm j}   , \quad 
J_{\rm j}=\rho_{\rm j}\Gamma_{\rm j}v_{\rm j}   ,
\end{eqnarray}
where
${\hat \gamma}$,
$V$, 
$A_{\rm j}$,
$T_{\rm j}^{01}$,
$J_{\rm j}$,
$\rho_{\rm j}$, and 
$\Gamma_{\rm j}$
are
the adiabatic index of the plasma in the cocoon,
the volume of the cocoon, 
the cross-sectional area,
the total energy flux, and 
rest mass flux,
rest mass density, and 
bulk Lorentz factor of the jet, 
respectively.
The term $V$ is evaluated as
$V=2(\pi/3){\cal R}^{2}LS^{3}$, where
$LS$ and $\cal R$ are the linear size 
of the cocoon along the jet axis
and the aspect ratio of the cocoon, respectively.
Here we denote physical quantities 
of the jet with the subscript j.
Throughout this work, 
we focus on a relativistic jet.
Correspondingly, the shocked plasma has
relativistic energy, thus we take $\hat \gamma=4/3$.
The $PdV$ work done by the cocoon against ICM is 
taken into account in the energy equation  Eq. (\ref{eq:pc}) 
following  I08.
For given $\rho_{\rm ICM}$,
we can dynamically estimate total pressures $P$
by measuring $LS$, ${\cal R}$, and
the head cross-sectional area of the cocoon.
Here the relations of 
$LS=\beta_{\rm hs}ct_{\rm age}$ and 
${\cal R}\equiv l_{\rm c}/LS<1$  hold
where 
$l_{\rm c}$ and $\beta_{\rm hs}c$
are the lateral size of the cocoon and
advance velocity of the hot spot, respectively.
Since ${\cal R}$ and $\beta_{\rm hs}$ have some 
uncertainties,
actual $P$ is bounded by maximum 
and minimum values  
\begin{eqnarray}\label{eq:prange}
P_{\rm min}\le P \le P_{\rm max}.
\end{eqnarray}
%
Thus we can obtain the total pressure of cocoon $P$, 
which includes the partial pressures of non-radiating particles.
The estimate of $P$ has actually been done by I08  
for some FR II sources and we adopt $P$ values
in I08 in this work.

\section{Pressure as a function of $\eta$}

In this section, we express $P$ as  
a sum of the partial pressures and 
represent it as a function of $\eta$ 
(we call this Equation of State, EOS)
for respective cases.

\subsection{Case (a)}

First, we examine the canonical case of two-temperature
thermal plasma.
Here we assume that 
$P_{-}^{\rm NT}=P_{+}^{\rm NT}=P_{p}^{\rm NT}=0$ 
and 
$n_{-}^{\rm NT}=n_{+}^{\rm NT}=n_{p}^{\rm NT}=0$.
The EOS in the cocoon
 filled with relativistic plasma is given by
\begin{eqnarray}\label{eq:EOS}
P&\approx &P^{\rm T}_{\pm}+P^{\rm T}_{p} \nonumber \\ 
 &=&(n_{-}^{\rm T}+n_{+}^{\rm T})kT_{\pm}+n_{p}^{\rm T}kT_{p}   ,
\end{eqnarray}
where 
$T_{\pm}$, and
$T_{p}$
are
the electron/positron temperature, and
proton temperature, respectively. 
Hereafter we adopt $T_{\pm}=T_{-}=T_{+}$ where 
$T_{-}$ and $T_{+}$ are temperatures of 
electrons and positrons, respectively.
Following Kino et al. (2007),
we can obtain $T_{\pm}$ and $T_{p}$ 
from Eqs. (\ref{eq:pc}), (\ref{eq:rho}), and (\ref{eq:EOS}):
\begin{eqnarray} \label{eq:T-two}
kT_{\pm} = \frac{\Gamma_{\rm j}m_{e}c^{2}}{4}, \quad
kT_{p} = \frac{\Gamma_{\rm j}m_{p}c^{2}}{4} ,
\end{eqnarray}
which are typically given by
$kT_{\pm} = 1.3~
\left(\frac{\Gamma_{\rm j}}{10}\right) ~ {\rm MeV}$, and 
$kT_{p} = 2.3 ~ 
\left(\frac{\Gamma_{\rm j}}{10}\right) ~ {\rm GeV}$.
Here we assume the limit of 
inefficient $e/p$-coupling i.e, 
protons and electrons are separately thermalized so that
$kT_{\pm}= (m_{e}/m_{p}) k T_{p}$
since  
plasma number densities in large scale jets are conservatively 
expected to be
too dilute to achieve efficient $e/p$-coupling
(e.g., Kino et al. 2007 and references therein).
The emission timescale is so long 
that radiative cooling is negligible.
It is worth noting that the geometrical
factors in Eqs. (\ref{eq:pc}) and (\ref{eq:rho}) are completely
canceled out and
$kT_{\pm}$ and $kT_{p}$ are  governed only by $\Gamma_{\rm j}$.

Inserting Eq. (\ref{eq:T-two}) into Eq. (\ref{eq:EOS}), 
we rewrite the total pressure in the cocoon $P$ as
\begin{eqnarray}\label{eq:P}
P(\eta)&=&
2.05 \times 10^{-6}~ 
n_{-}^{\rm T}\left[(2-\eta)+\eta\frac{m_{p}}{m_{e}}\right]
\left(\frac{\Gamma_{\rm j}}{10}\right)
 ~\rm erg \ cm^{-3} , \nonumber \\ 
\end{eqnarray}
where
the first term and second term in the square bracket
correspond to the partial pressure of pairs and protons,
respectively.

\subsection{Case (b)}

As an opposite extreme to case (a),
here we consider the case of one-temperature plasma. 
In this example, 
some of the proton energy is somehow
transferred to electrons/positrons 
to achieve an efficient $e/p$-coupling.
Then hotter electrons/positrons and 
colder protons are produced.
From the condition $kT_{\pm}=kT_{p}$,
and Eqs. (\ref{eq:pc}) and (\ref{eq:rho}), we obtain
\begin{eqnarray}
kT_{\pm}
=kT_{p}
=\frac{\Gamma_{\rm j}m_{e}c^{2}}{8} 
\left[(2-\eta)+\eta \frac{m_{p}}{m_{e}}\right] .
\end{eqnarray}
In this case,  each population (i.e., $p/e^{-}/e^{+}$)
has the same kinetic energy.
The total pressure is given by
Eq. ($\ref{eq:P}$) the same as case (a).
The essential difference from  case (a) 
is that $kT_{\pm}$ in case (b) 
is much higher than the one in case (a).


\subsection{Case (c)}

For comparison with the canonical case (a),
we examine case (c) when the cocoon 
pressure is dominated by non-thermal particles.
Case (c) concerns when
the spectral indices of non-thermal particle energy
distributions satisfy $s_{p}=s_{e}>2$
as some theoretical
work on relativistic shocks suggests  
(e.g., Bednarz and Ostrowski 1998;
Kirk et al. 2000;
Achterberg et al. 2001;
Spitkovsky 2008;
Sironi and Spitkovsky 2011) 
and as
the radio lobes of Cygnus A show $s_{e}>2$ 
(e.g., Carilli et al. 1991; Yaji et al. 2010).
In this case, electrons and protons with the 
lowest energies are the main carriers of energy.
Then, the evaluation of partial pressures of 
non-thermal plasma 
is basically the same as in case (a)
when we replace 
$kT_{\pm}$ with $\gamma_{\pm,\rm min}m_{e}c^{2}$
and  
$kT_{p}$ with $\gamma_{p,\rm min}m_{p}c^{2}$.
Then $P$ is given by
\begin{eqnarray}
P(\eta)=
\frac{\Gamma_{\rm j}n_{-}^{\rm NT}m_{e}c^{2}}{3}
\frac{s_{e}-1}{s_{e}-2}\left[(2-\eta)+\eta \frac{m_{p}}{m_{e}}\right]  .
\end{eqnarray}
From this, it is clear that
we can appropriately evaluate  $\eta$ for the case (c) 
by replacing $n_{-}^{\rm T}$ to $n_{-}^{\rm NT}$
in the same way as case (a).

%

\subsection{Case (d)}

Here we examine the pressure of non-thermal electrons 
when they follow a broken power law spectrum Eq. (\ref{eq:ne-d}). 
Stawarz et al. (2007) indicated
$\gamma_{\pm,\rm crit}\sim m_{p}/m_{e}$
for the hot spots in Cygnus A.
The energy of the electron component is governed by those 
with break energy, while the number is dominated 
by those with lowest energies.
Since $s_{p}>2$ is satisfied, 
lowest-energy protons carry the most energy.
Therefore, the total pressure $P$ is expressed as
\begin{eqnarray}\label{eq:Pcase-e}
P(\eta)=
\frac{\Gamma_{\rm j}n_{-}^{\rm NT}m_{e}c^{2}}{3}
\left[
\frac{s_{e,1}-1}{-s_{e,1}+2}
A_{\pm}(2-\eta)+
\frac{s_{p}-1}{s_{p}-2}
\eta \frac{m_{p}}{m_{e}}\right] , \nonumber \\
\end{eqnarray}
where 
$A_{\pm}= (\gamma_{\pm,\rm crit}
/\gamma_{\pm,\rm min})^{-s_{e,1}+2}$.
Thus $\eta$ can be evaluated
when we replace  $n_{-}^{\rm T}$ to $n_{-}^{\rm NT}$
and include factor $A_{\pm}$.

\section{Testing plasma composition}

We explain the method for 
constraining plasma composition of AGN jets
for thermal plasma cases (a) and (b) in 5.1, 5.2, and 5.3.
The application to non-thermal plasma cases (c) and (d) 
can be readily 
understood and is explained in 5.4.

\subsection{Characteristic pressures}

\begin{figure} 
\includegraphics[width=10cm]{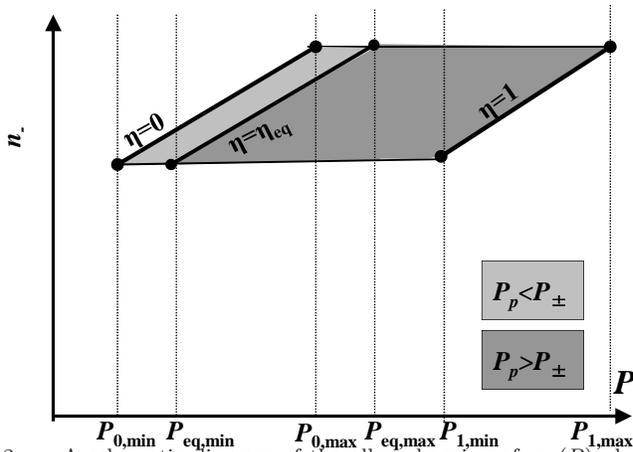}
\caption
{
A schematic diagram of the allowed region of $n_{-}(P)$
plotted versus the cocoon pressure $P$
for given $\Gamma_{\rm j}$.
They are limited by  $ n_{-,\rm min}\le n_{-} \le  n_{-,\rm max}$
and $0\le \eta \le 1$.
The region of
$e^{\pm}$-supported cocoon ($P_{\pm}>P_{p}$) is colored in 
light gray while the region of
proton-supported cocoon ($P_{p}>P_{\pm}$) is colored in
dark gray. }
\label{fig:npelectron}
\end{figure}

Firstly we define characteristic pressures which
divide the number-density/pressure plane
into several regions as shown in Fig. \ref{fig:npelectron}.
As a preparation,
here we define $\eta_{\rm eq}$
\begin{eqnarray}
\eta_{\rm eq}&\equiv& \frac{2}{m_{p}/m_{e}-1}=
1.1\times 10^{-3} \quad (P_{\pm}=P_{p}) .
\end{eqnarray}
The partial pressure of proton-associated electrons is
implicitly neglected since it is subdominant 
in the case of inefficient $e/p$-coupling.
The line with $n_{-}=1\times 10^{3}n_{\rm p}$
divides the pair-supported and proton-supported cocoon
in the limit  of inefficient $e/p$-coupling plasma.
By definition,
the cocoon with $\eta>\eta_{\rm eq}$ is proton-supported 
(dark gray region in Fig. \ref{fig:npelectron})
while the cocoon with $\eta<\eta_{\rm eq}$ is 
pair-supported one (light gray region in Fig. \ref{fig:npelectron}).
When $n_{-}$ is bounded by $n_{-, \rm min}$ 
and $n_{-, \rm max}$ as argued in the next subsection,
the allowed region
of $n_{-}$ is segmented by some
characteristic pressures by
the characteristic values of $n_{-}$ and $\eta$,
i.e., $n_{-,\rm min}$, $n_{-,\rm max}$, $\eta=0$,
 $\eta=\eta_{\rm eq}$, and  $\eta=1$.
Here, we define six characteristic pressures as follows;
\begin{eqnarray}
P(\eta=0;n_{-}=n_{-,\rm min})            &\equiv& P_{0,\rm min},  \nonumber \\ 
P(\eta=\eta_{\rm eq};n_{-}=n_{-,\rm min})&\equiv& P_{\rm eq,min},   \nonumber \\  
P(\eta=0;n_{-}=n_{-,\rm max})            &\equiv& P_{0,\rm max},  \nonumber \\ 
P(\eta=\eta_{\rm eq};n_{-}=n_{-,\rm max})&\equiv& P_{\rm eq,max},\nonumber \\ 
P(\eta=1;n_{-}=n_{-,\rm min})            &\equiv& P_{1,\rm min},   \nonumber \\ 
P(\eta=1;n_{-}=n_{-,\rm max})            &\equiv& P_{1,\rm max}  . 
\end{eqnarray}
Then, by definition, we have the following relations
\begin{eqnarray}
P_{0,\rm min}:P_{\rm eq,min} :
P_{0,\rm max}:P_{\rm eq,max} 
:P_{1,\rm min}
:P_{1,\rm max} \nonumber \\
= 1:2
:\frac{n_{-,\rm max}}{n_{-,\rm min}}
:2\frac{n_{-,\rm max}}{n_{-,\rm min}}
:\frac{m_{p}}{m_{e}}
:\frac{m_{p}}{m_{e}}\frac{n_{-,\rm max}}{n_{-,\rm min}}  ,
\end{eqnarray}
where we approximate $2-\eta_{\rm eq}\approx 2$.
To evaluate these pressures, 
we estimate $n_{-,\rm min}$ and
$n_{-,\rm max}$ in the next subsection.

\subsection{Estimation of $n_{-}$}

Here we  constrain  the
number density of electrons in the cocoon ($n_{-}$).
We denote the lower and upper limits of $n_{-}$ as 
$n_{-,\rm min}$ and  $n_{-,\rm max}$, respectively.
The values of $n_{-,\rm min}$ and  $n_{-,\rm max}$ are
independently constrained and we show them below.

%

\subsubsection{Lower limit of $n_{-}$}

Here we estimate the lower limit of $n_{-}$
and examine the case when 
the number density of thermal electrons 
is larger than that of non-thermal electrons
$n_{-}^{\rm T}\ge n_{-}^{\rm NT}$, since
non-thermal electrons are partially injected 
from the background thermal electrons.
(Later, 
the extreme cases of $n_{-}^{\rm T}\le n_{-}^{\rm NT}$
will also be discussed, being identical 
to cases (c) and (d)).
Since the shocked plasma at hot spots expands 
sideways and is injected into the cocoon,
we can estimate $n_{-}^{\rm NT}$
by using $n_{\rm hs}^{\rm NT}$ where
$n_{\rm hs}^{\rm NT}$ is the number density of non-thermal
electrons in a hot spot.
We stress that $n_{\rm hs}^{\rm NT}$
is well constrained by observed non-thermal emissions of 
hot spots for FR II sources 
(e.g., Harris and  Krawczynski 2006 for review).
By connecting the 
number density from the jet to the cocoon
based on 
Eq.~(\ref{eq:rho}) and shock conditions along the jet axis
shown in Kino and Takahara (2004)( hereafter KT04), 
we obtain
\begin{eqnarray}
n_{-,\rm min}&=&
\frac{n^{\rm NT}_{\rm hs}  A_{\rm j} LS}
{2V\beta_{\rm hs}} .
\end{eqnarray}
%
In general, number density of
non-thermal electrons with  power law distribution
$n^{\rm NT}_{\rm hs}
\propto 
\int^{\gamma_{\rm hs,max}}
_{\gamma_{\rm hs,min}}\gamma_{\rm hs}^{-s_{\rm hs}} d\gamma_{\rm hs}$
can be given by
\begin{eqnarray}\label{eq:nhs}
n^{\rm NT}_{\rm hs}\propto   \gamma_{\rm hs,min}^{-s_{\rm hs}+1}  .
\end{eqnarray}
We assume the standard value of $s_{\rm hs}\approx 2$ and 
$\gamma_{\rm hs,min}\approx \Gamma_{\rm j}$.




\subsubsection{Upper limit of $n_{-}$}

The upper limit of $n_{-}$ can be
constrained 
by the absence of thermal bremsstrahlung
from hot electrons in the cocoon/lobes viewed in $X$-ray
observations (Wilson et al. 2000, 2006).
The observed  $X$-ray emissions associated with
radio lobes are non-thermal emissions and 
there is no evidence for thermal $X$-ray
emission from coccons/lobes
(Harris and Krawczynski 2006 for review).
From this, we can safely use the condition of 
$L_{{\rm X, obs}}>L_{\rm brem}(n_{-}^{\rm T},T_{\pm})$ 
where 
$L_{\rm brem}/V=
\alpha_{\rm f}r_{e}^{2}m_{e}c^{3}(n_{-}^{\rm T})^{2}F_{\pm}(\Theta_{\pm})
~{\rm erg~s^{-1}~cm^{-3}}$, 
$F_{\pm}(\Theta_{\pm})=
48\Theta_{\pm}(\ln 1.1\Theta_{\pm}+5/4)$, and 
$\Theta_{\pm}=kT_{\pm}/m_{e}c^{2}$,
for bremsstrahlung at
relativistic temperature (Eq. (22) in Svensson 1982)
and $\alpha_{\rm f}$ and $r_{e}$ are the fine structure constant
and the classical electron radius, respectively.
From this, we obtain the maximum  $n_{-}$ as follows:
\begin{eqnarray}\label{eq:Lx}
n_{-,\rm max} =
\left(\frac{L_{\rm brem}}
{V\alpha_{\rm f}r_{e}^{2}m_{e}c^{3}
F_{\pm}(\Theta_{\pm})}\right)^{1/2}    .
\end{eqnarray}
%

It is worth commenting on
the availability of constraining the upper limit of
$n_{-}$ by the analysis of the internal depolarization 
of the radio lobes. 
Relativistic plasma
makes a smaller contribution to 
Faraday rotations 
since electron inertia 
increases for the relativistic regime 
and it suppresses rotations of polarization angle
(e.g., 
Ichimaru 1973;
Melrose et al. 1997;
Quataert and Gruzinov 2000;
Huang and Shcherbakov 2011).
Therefore, it is not effective to use
the constraint by RM in the present work.

\subsection{Estimation of $n_{p}$}

\begin{figure} 
\includegraphics[width=10cm]{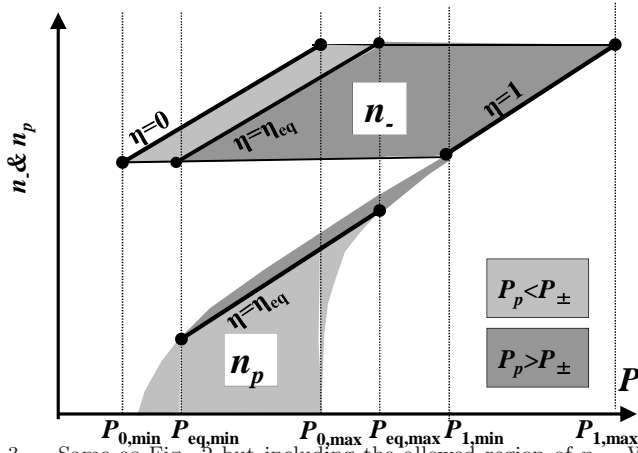}
\caption
{Same as
Fig. \ref{fig:npelectron} but 
including the allowed region of  $n_{p}$. 
When $\eta=1$,  $n_{-}=n_{p}$ holds by definition.
When $\eta<1$, positron mixture is required 
by the charge neutrality condition of $n_{-}=n_{p}+n_{+}$.
The plane is divided into 5 regions
by characteristic pressures.
The actual allowed region is further
limited within $P_{\rm min}\le P \le P_{\rm max}$
by the consideration of cocoon calorimetry.}
\label{fig:np-cartoon}
\end{figure}

Once $n_{-}$ is estimated, 
the proton number density  $n_{p}$
can be determined as 
\begin{eqnarray}\label{eq:nproton}
n_{p} =\eta n_{-}  ,
\end{eqnarray}
by definition.
Here, of course, the conditions of $0\le \eta \le 1$ 
and $n_{-,\rm min}\le n_{-}\le n_{-,\rm max}$ are imposed.
In Fig. \ref{fig:np-cartoon}, the allowed region
of $n_{p}$ is added to that of
 $n_{-}$ shown in Fig. \ref{fig:npelectron}.
In the same way as Fig. \ref{fig:npelectron}, the plane is divided
into 5 regions.

Finally, the allowed regions of 
$n_{p}$ and $n_{-}$ can be 
obtained by adjoining  the range of $P$.
The allowed regions drawn in Fig. \ref{fig:np-cartoon}
are  bounded by Eq. (\ref{eq:prange}).
Thus, we can obtain the definitive
allowed regions of $n_{p}$ and $n_{-}$.

\subsection{Application to cases (c) and (d)}

In 5.1, 5.2 and 5.3,
we consider physical quantities
associated with thermal plasma 
in  cases (a) and (b).
But those can also be applied to non-thermal plasma
by the proper replacements of number densities and 
average energies of particles.
With regard to average energies, we 
have already explained the replacements
in the previous section.
As for $n_{-,\rm min}$, the estimate shown in 5.2.1. 
can be applied both for thermal and non-thermal plasma.
As for $n_{-,\rm max}$, the estimate shown in 5.2.2. 
can be applied only for thermal plasma.
So, we do not use $n_{-,\rm max}$ for the cases (c) and (d).
Thus
we can properly estimate $\eta$ for cases (c) and (d).

\section{Application to Cygnus A}

Here we apply the above method to Cygnus A ($z=0.0562$)
which is one of the best studied FR II radio galaxies
(e.g., Carilli and Barthel 1996; Steenbrrugge et al. 2008, 2010;
Yaji et al. 2010).
The physical quantities of Cygnus A
have been well constrained by previous work.
To constrain the
real values of  $P$ and  $n_{-}$, 
we carefully evaluate 
${\cal R}$,  
$\beta_{\rm hs}$, and 
$\Gamma_{\rm j}$.
The term ${\cal R}$ has an effect on 
$n_{-}$ via a cocoon volume $V$. 
The term $\Gamma_{\rm j}$ is directly 
proportional to $P$.
The term $\beta_{\rm hs}$ controls the 
source age $t_{\rm age}$ which governs the injection rates of 
mass and energy into the cocoon.
These are summarized in 6.1.
The resultant 
allowed region of $n_{-}$ and $n_{p}$ is summarized in 6.2.

\subsection{Viable ranges of physical quantities}

We show adopted conditions of the model parameters 
for deriving the above results.
We fix the cross section area of the jet as
$A_{\rm j}=\pi R_{\rm hs}^{2}=\pi (2~{\rm kpc})^{2}$ 
(Wilson et al. 2000) and
the number density of ICM just ahead of the 
hot spot as
$n_{\rm ICM}=0.5 \times10^{-2} ~{\rm cm^{-3}}$ 
(the shell No. 6 in Table 5 in Smith et al. 2002).

\begin{itemize}

\item

{\em Cocoon morphology ${\cal R}$.} 

From images of the Cygnus A cocoon,
we can directly constrain ${\cal R}$. 
The upper limit ${\cal R}\approx0.5$ is determined by the Chandra
X-ray image (Wilson et al. 2000, 2006; Yaji et al. 2010).
The lower limit ${\cal R}\approx0.25$ is directly measured by
the 330 MHz VLA image (see also Carilli et al. 1991; Lazio et al. 2006).
Therefore we set
\begin{eqnarray}
0.25\le {\cal R} \le 0.5 , \nonumber 
\end{eqnarray}
in the present work.

\item

{\em Cocoon head velocity $\beta_{\rm hs}$.} 

Cocoon head velocity which equals the 
hot spot advance velocity ($\beta_{\rm hs}$) 
is well constrained by the  synchrotron aging method.
The estimated $\beta_{\rm hs}$ has some uncertainty
due to the uncertainty of magnetic-field strength in the
cocoon.
From the result of synchrotron aging diagnosis 
in Carilli et al. (1991),
we adopt the allowed range of $\beta_{\rm hs}$ as
\begin{eqnarray}
0.01 \le \beta_{\rm hs} \le 0.06. \nonumber 
\end{eqnarray}
We emphasize that sufficiently large uncertainty is taken
into account here. 
The adopted value of $ \beta_{\rm hs}$  is 
quite typical for hot spots in FR II radio galaxies 
(e. g., Scheuer 1995). 

\item

{\em Lorentz factor of the jet $\Gamma_{\rm j}$.} 

It is difficult to determine the 
true velocity of the jet.
At least we may say that 
apparent velocity of blobs obtained by 
VLBI observations show a minimum velocity of underlying flow. 
A fast apparent motion of a blob at the jet base 
$(0.56\pm 0.28)~c$ has been 
reported by VLBI observations (Bach et al. 2003).
Furthermore, suggestions of superluminal 
motion were made
(Krichbaum et al. 1998; Bach et al. 2002)
although they had not been clearly confirmed.
On VLA scale, a clear asymmetry in brightness distribution 
of a kpc-scale jet due to a relativistic motion is seen
(Perley et al. 1984). 
Therefore, overall radio observations seem to
indicate relativistic motion.
Bearing this in mind, 
we assume that the jet is relativistic
and the four-velocity of the jet $ \Gamma_{\rm j}\beta_{\rm j}$
is set as
\begin{eqnarray}
1 \le \Gamma_{\rm j}\beta_{\rm j} \le 30.  \nonumber 
\end{eqnarray}
%
Here the upper limit is assumed as $\Gamma_{\rm j}\approx 30$
based on the statistical study of radio jets 
of MOJAVE sources
(Lister et al. 2001, 2009; Kellermann et al. 2004).
%

\item

{\em Cocoon pressure $P$.} 

Using the value of
 $V=1\times 10^{70}{\cal R}^{2}~{\rm cm^{3}}$, 
we can estimate the total pressure $P$ as
\begin{eqnarray}\label{eq:P-cygnusA}
8\times 10^{-11}~{\rm erg \ cm^{-3}} 
\le P \le
4\times 10^{-9}~{\rm erg \ cm^{-3}}.
\end{eqnarray}
The lower limit equals 
the ICM pressure $8\times 10^{-11}~{\rm erg \ cm^{-3}} $
measured by Arnaud et al. (1984)
to satisfy the over-pressured cocoon condition. 
Although the upper limit of $P$ is basically adopted from I08,
the value $4\times 10^{-9}~{\rm erg \ cm^{-3}}$
is 4 times larger than the original estimate in I08.
This is due to the change in minimum value of ${\cal R}$
from 0.5 to 0.25 based on VLA's 0.3 GHz image.
It should be stressed that our adoption of 
the allowed range of $P$ is sufficiently wide compared with 
all of the previous work (e.g, Carilli 1998 for review).
Note that Yaji et al. (2010) estimates
that $P_{-}^{\rm NT}$ in the radio lobes as
$P_{-}^{\rm NT}\approx (1-2)\times 10^{-9}~{\rm erg~cm^{-3}}$
for $\gamma_{\pm}\approx 1$ which 
causes $P^{\rm NT}_{-}>P_{\rm min}$.
So, if $P$ completely equals the radio lobe pressure,
then the range $P_{\rm min}\le P<P^{\rm NT}_{-}$ is 
excluded and the allowed $P$ range becomes narrower.
The allowed example with
$P_{\rm min}\le P^{\rm NT}_{-} \le P\le P_{\rm max}$ is 
involved in cases (c) and (d).

\item
{\em Non-thermal electron number density $n^{\rm NT}_{-,\rm hs}$.} 

The lower limit $n_{-,\rm min}$  largely depends on 
$n^{\rm NT}_{\rm hs}$.
For $s_{\rm hs}=2$, 
the number density of non-thermal electrons in the 
hot spot can be obtained from
\begin{eqnarray}\label{eq:nNT}
n^{\rm NT}_{\rm hs}
\approx 1 \times 10^{-3}
\left(\frac{\gamma_{\rm hs,min}}{10}\right)^{-1} ~{\rm cm^{-3}}   ,
\end{eqnarray}
via detailed comparisons of
the SSC model with the observed broadband spectrum 
(Wilson et al. 2000; KT04; Stawarz et al. 2007)
where $\gamma_{\rm hs,min}\approx \Gamma_{\rm j}$.
We stress that these three independent papers 
derive similar values of  $n^{\rm NT}_{\rm hs}$
although Stawarz et al. 2007)
adopts the different 
electron-distribution function shown in Eq. (\ref{eq:ne-d}).
Furthermore, we note the importance of low-frequency radio 
spectra since it affects the estimate of $n^{\rm NT}_{\rm hs}$. 
Regarding low-frequency radio observation, 
we briefly comment on the work of Lazio et al. (2006). 
They indicated  spectral flattening and turnover 
at $\sim 100~{\rm MHz}$.
However it seems difficult to determine these accurately 
because the spot sizes are smaller than the VLA beam sizes
at the above frequencies.
The LOw Frequency ARray 
(LOFAR) (http://www.lofar.org/) and
Square Kilometer Array (SKA) 
(http://www.skatelescope.org/)
will, in future, tell us the real turnover frequency
with sufficiently high resolution.

\item
{\em Thermal electron number density $n_{-}^{\rm T}$.}

Here we comment on the difficulty of constraining
$n_{-}^{\rm T}$. 
We use the absence of bremsstrahlung emission. 
The $X$-ray observations for Cygnus A show the flux
upper limit as
$\sim 1 \times 10^{-13}~{\rm erg~s^{-1}~cm^{-2}}$
(e.g., Smith et al. 2002).

As already mentioned, the constraint 
from the intrinsic RM is not available, 
because plasma temperature is relativistic 
in the present work.
Even worse, Cygnus A is known for
its unusually large RM values
and thus it is not a good example from which 
to argue the intrinsic depolarization 
(Dreher et al. 1987; Garrington and Conway 1991).
No evidence for intrinsic depolarization
between 5 and 15 GHz is found 
and the origin of the large RM is thought to be
the external bow shock 
which surrounds the radio lobes 
(Dreher et al. 1987; Carilli et al. 1988).
Hence it is not appropriate to use
the constraint from RM for Cygnus A.

\end{itemize}

\subsection{Results}

Below we show resultant allowed region of 
$n_{-}$ and $n_{p}$ for 
cases (a), (b), (c) and (d).

\subsubsection{Case (a)}

Considering
the uncertainties of 
$\Gamma_{\rm j}\beta_{\rm j}$ and $\beta_{\rm hs}$,
we examine two limiting cases with
$\Gamma_{\rm j}\beta_{\rm j}=1$ 
and $\beta_{\rm hs}=0.01$ 
being a High-$n$ case, and 
that with $\Gamma_{\rm j}\beta_{\rm j}=30$ 
and $\beta_{\rm hs}=0.06$ being a Low-$n$ case.
For the High-$n$ case, $n_{-}$ is about two orders 
of magnitude larger than that of the Low-$n$ case.

In Fig. \ref{fig:npH}, we show
the allowed region of $n_{-}$ and $n_{p}$ 
for the High-$n$ case.
First of all, we find that 
$n_{-}>n_{p}$ always holds 
and this satisfies
$\eta\sim 10^{-2}$ at $P=P_{\rm max}$.
This implies that positron mixture is inevitable.
In other words, $P_{1,\rm min}$ is 
much larger than $P_{\rm max}$ 
obtained by the Cygnus A cocoon calorimetry.
(If we are force to make $P_{1,\rm min}$ smaller, 
then $\gamma_{\rm min}$ becomes larger and
such a case coincides with (b).)
The allowed regions of $n_{-}$ and $n_{p}$
are further divided by two regions.
The pair of light-gray regions show the one in which
$P_{\pm}>P_{p}$ is satisfied. 
On the contrary, the pair of dark-gray 
regions display the one in which
$P_{\pm}<P_{p}$ holds. 
Interestingly, we find that  the 
regions of $P_{p}<P_{\pm}$ and $P_{p}>P_{\pm}$ are both
wide in the range of allowed $P$. 
Only in the range of
$P\sim (3-6)\times 10^{-10}~{\rm erg cm^{-3}}$,
the pair dominance $P_{p}<P_{\pm}$ alone is permitted
in the High-$n$ case.

Fig. \ref{fig:npL} displays the result for 
the Low-$n$ case.
Similar to the High-$n$ case,
$n_{-}>n_{p}$ always holds 
and they satisfy $\eta\sim 10^{-1}$ at $P=P_{\rm max}$.
Due to the decrease in $n_{-,\rm min}$,
the number densities in allowed regions
are  about two orders of magnitude smaller
than that for the High-$n$ case shown in Fig.  \ref{fig:npH}.
Correspondingly, 
$P_{0,\rm min}$, 
$P_{\rm eq, min}$, and
$P_{1,\rm min}$  decrease.
Since $P_{\rm eq,max}<P_{\rm max}$ is still satisfied,
both of the regions with $P_{p}<P_{\pm}$ and  that with 
$P_{p}>P_{\pm}$ are allowed in this case. 
In other words, the Low-$n$ case also draws the same conclusion 
with the High-$n$ case qualitatively.
Quantitatively, 
the upper limit of $n_{p}$ becomes larger 
when $n_{-,\rm min}$ becomes smaller
and correspondingly 
the maximum $\eta$ achieved at $P_{\rm max}$
becomes larger by a factor of $\sim 10$ 
than that for the High-$n$ case.

In summarizing case (a), we find that 
$\eta<1$ always holds in the allowed range of $P$.
In other words, this indicates the existence of $e^{\pm}$ pairs
in the cocoon. 
We find that (i)
$e^{\pm}$ pair is dominant in terms of number density, and
(ii) both the ``pair-supported cocoon (i.e., $P_{\pm} >P_{p}$)'' 
and the ``proton-supported one (i.e, $P_{\pm} <P_{p}$)'' 
are allowed.
The pair-supported cocoon is different 
from the previously suggested one in which protons are 
dynamically dominated (e.g., De Young 2006).

\subsubsection{Case (b)}

\begin{figure}
\includegraphics[width=10cm]{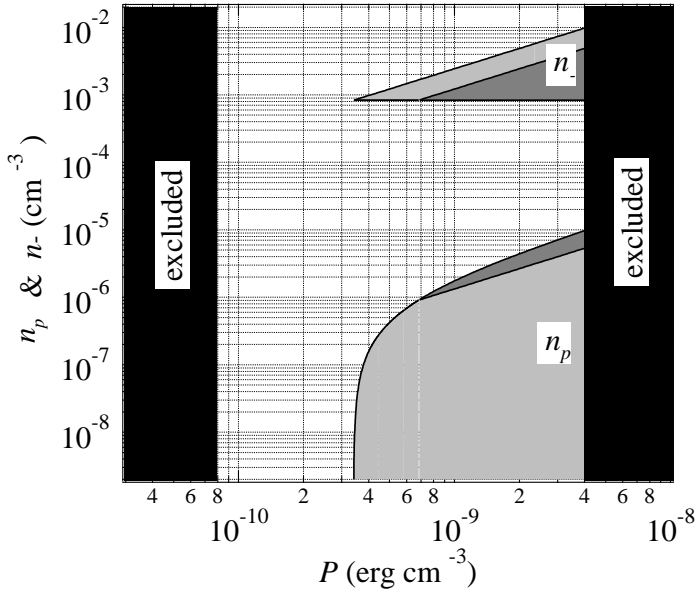}
\caption
{Allowed regions of $n_{-}$ and $n_{p}$ for Cygnus A
with $\Gamma_{\rm j}\beta_{\rm j}=1$ 
and $\beta_{\rm hs}=0.01$
(we call this the High-$n$ case).
The region within 
$8\times 10^{-11}~{\rm erg \ cm^{-3}} 
\le P \le
4\times 10^{-9}~{\rm erg \ cm^{-3}}$ 
shown here is the one allowed for Cygnus A.
As explained in Fig. \ref{fig:np-cartoon},
the region in which $P_{\pm}>P_{p}$ holds is colored in 
light gray while the region where $P_{\pm}>P_{p}$ is satisfied 
is colored in dark gray. 
It is found that 
$e^{\pm}$ pairs always dominate in terms of number density
but either  ``pair-supported cocoon (i.e., $P_{\pm} >P_{p}$)'' 
or ``proton-supported one (i.e, $P_{\pm} <P_{p}$)'' is possible.}
\label{fig:npH}
\end{figure}

\begin{figure} 
\includegraphics[width=10cm]{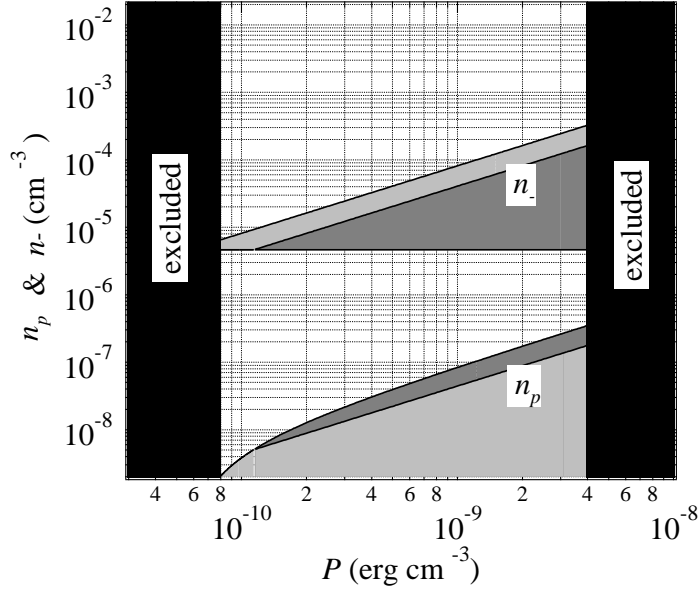}
\caption
{
Same as Fig. \ref{fig:npL} 
but with $\Gamma_{\rm j}=30$ and $\beta_{\rm hs}=0.06$
(we call this the Low-$n$ case).
Although the allowed regions of  
$n_{-}$  and $n_{p}$ are 
about two orders of magnitude smaller
than the ones in  Fig. \ref{fig:npH} (High-$n$ case),
the Low-$n$ case also draws the same conclusion 
as the High-$n$ case.}
\label{fig:npL}
\end{figure}

For Cygnus A, we face a difficulty 
of realizing one-temperature plasma.
First, let us consider the case of 
same $n_{-,\rm min}$ as 
in Figs. \ref{fig:npH} and \ref{fig:npL}. 
Then all of these thermal electrons should be heated up to 
$kT_{\pm}\sim 10^{4}m_{e}c^{2}$ 
and injected into the lobes in the case (b).
In the radio lobes, Yaji et al. (2010) evaluates 
the number density of non-thermal electrons as 
$\sim 10^{-7}~{\rm cm^{-3}}$ 
at $\gamma_{-}\sim 10^{4}$.
So, if we allow the existence  of thermal plasma 
with the same $n_{-,\rm min}$ 
in Figs. \ref{fig:npH} and \ref{fig:npL}
but with $kT_{\pm}=kT_{p}\sim 10^{4}m_{e}c^{2}$,
a big thermal bump at $\sim 10^{9}~{\rm Hz}$ should appear.
However there is no such bump in the observed 
spectra of the radio lobes. Therefore, we can exclude 
the case of the same $n_{-,\rm min}$ 
with $kT_{\pm}=kT_{p}\sim 10^{4}m_{e}c^{2}$.

Next, we consider  smaller $n_{-,\rm min}$.
Using the relation
$n_{-,\rm min}\propto \gamma_{\rm hs,min}^{-1}$ 
in Eq.~(\ref{eq:nhs}),
the increase in $\gamma_{\rm hs,min}$ 
leads to the decrease in $n_{-,\rm min}$ 
in Figs. \ref{fig:npH} and  \ref{fig:npL};
basically,
$\gamma_{\rm hs,min}\sim 10^{4}$ is required at the hot spot
(e.g., 
Harris et al. 2000;
Hardcastle, Birkinshaw, and Worral 2001;
Blundell et al. 2006; Godfrey et al. 2009).
However, in the case of Cygnus A,
the model spectra of the hot 
spots with $\gamma_{\rm hs,min}\ge 2000$ 
conflict with the observed ones (KT04).
Therefore, case~(b) is not likely for Cygnus A.

\subsubsection{Case (c)}

Let us consider the case of
dominant non-thermal pressures and a separate
acceleration of electrons and protons with 
a steep  power law spectrum.
This is almost identical to (a).
A slight difference between this case and (a)
is the evaluation of $n_{-,\rm min}$.
Since  non-thermal pairs are dominated 
in this case, the allowed region would be limited around
$n_{-}\approx n_{-,\rm min}$ in 
Figs. \ref{fig:npH} and \ref{fig:npL}.

%

\subsubsection{Case (d)}

Let us consider case (d).
The factor 
$A_{\pm}= (\gamma_{\pm,\rm crit}
/\gamma_{\pm,\rm min})^{-s+2}$ in 
Eq. (\ref{eq:Pcase-e}) is the only element
to change the result from (a).
Since $\gamma_{\rm crit,\pm}\sim m_{p}/m_{e}$
is suggested by Stawarz et al. (2007),
we can estimate $A_{\pm}$ as  
$A_{\pm} \approx 14(\Gamma_{\rm j}/10)^{0.5}$ 
for $s_{e,1}= 1.5$.
Therefore,
a difference between this case and (a)
is the larger $P_{\pm}$ by a factor of $A_{\pm}$.
Although the spectral break may be suggested from 
radio observations for  case (d),
$n_{-}^{\rm NT}$ is dominated by electrons 
at a break energy $\gamma_{\rm crit,\pm}m_{e}c^{2}$
and proton energies are not  entirely transported 
to electrons.
Therefore, results of (d) are expected 
to be intermediate between cases (a) and (b).

%

%

\section{Summary and discussions}

In this work, we  propose a new method for 
testing plasma composition of AGN jets
by using the cocoon dynamics.
In particular, 
we properly evaluate partial pressures
of protons and $e^{\pm}$ pairs.
The point of the method is 
that $n_{p}$ and $P_{p}$ can be constrained 
by considering the global conservations of kinetic energy,
mass, and momentum of shocked plasma in the cocoon.
Regarding particle distribution functions in the cocoon,
it is hard to determine  them uniquely because of 
sparseness of observational data.
Therefore, we examine four typical cases in this work.
Cases (a), (b), (c) and (d) respectively
present 
two-temperature thermal plasma, 
one-temperature thermal plasma,
non-thermal plasma with their spectral indices harder than two, and
non-thermal plasma with a broken power-law electron spectrum.

The three significant advantages of the present work
compared with previous work are summarized 
as follows;
\begin{enumerate}
\item
$P$ estimate is based on global 
cocoon dynamics. 
Since it is beaming-independent calorimetry of the 
true amount of energy released by the jet, 
the estimate of $P$ from cocoon dynamics
has fewer uncertainties compared with blazar studies.

\item
We focus on powerful FR II sources. 
Relativistic hydrodynamic simulations tell us that
FR II sources have less entrainment phenomena than FR I sources.
Therefore, FR IIs are better for 
testing genuine plasma composition of AGN jets.

\item
We properly deal with the partial pressure of 
thermal electrons/positrons $P_{\pm}^{\rm T}$. 
Although $P_{\pm}^{\rm T}$ is a critically important finite quantity, 
most prior efforts assume $P_{\pm}^{\rm T}=0$ merely for simplicity.

\end{enumerate}
Applying the method  
to the best studied FR II source Cygnus A, 
we draw the following conclusions
which primarily indicate the existence 
of numerous $e^{\pm}$ pairs in the cocoon of Cygnus A.
\begin{itemize}

\item

Cases (a), (c) and (d),
in which the average energy of 
electrons and positrons is significantly
lower than that of protons
($\eta <10^{-1}$ for Low-$n$ case;
$\eta <10^{-2}$ for High-$n$ case), are allowed 
without violating the observational constraints.
The results in (a) and (c) are 
almost the same, except that
the lowest energy electrons are thermal ones 
and non-thermal ones for (a) and (c), 
respectively.
Cases (a) and (d) also show similar results 
but for a larger $P_{\pm}$ in (d)
by a factor of $\sim 14$ than the one in (a).

\item

We can rule out case (b) in which electrons 
and positrons are heated up to the 
proton temperature of $\sim 10^{4}m_{p}c^{2}$.
Because there is no thermal bump
due to the hot thermal plasma.

\item

For (a), (c) and (d),
we find that the number density of $e^{\pm}$
is larger than $n_{p}$ in any allowed $P$ and
the obtained $n_{+}$ is always
more than 10 times larger than $n_{p}$.
We conclude that pure $e/p$ plasma is excluded
and $e^{\pm}$-proton mixture composition  
is achieved in the Cygnus A jet.
Therefore, further studies on the
$e^{\pm}$ pair loading problem 
extending previous ones
(e. g., 
Blandford \& Levinson 1995; 
Li \& Liang 1996; Thompson 1997; 
Beloborodov 1999;
Yamasaki, Takahara \& Kusunose 1999) 
will be more important and
the study of its bulk acceleration of  $e^{\pm}$ outflow
(Iwamoto and Takahara 2002, 2004; 
Asano and Takahara 2007, 2009)
will also be highly motivated.

\item

We find that  both   
$e/p$ plasma and $e^{\pm}$ pair pressure supported scenarios
are permitted
within the limit of current observational constraints.
We quantitatively show the allowed regions of
$P_{p}>P_{\pm}$ and $P_{p}<P_{\pm}$ by our new method
(see  Figs. \ref{fig:npH} and \ref{fig:npL}).


\end{itemize}

Lastly we add  a brief comment on $P_{p}^{\rm NT}$.
Recently Atoyan and Dermer (2008) has suggested
the possibility of a secondary emission 
induced by high-energy protons at Cygnus A. 
The luminosity of the secondary emission 
depends on $P_{p}^{\rm NT}$.
If the emission is detected in the future, 
it will provide us a new direct
constraint on $P_{p}^{\rm NT}$.
It could also give us a new constraint
on cosmic-ray propagations influenced by
the galactic magnetic field (Dermer et al. 2009).

%

\section*{Acknowledgments}
We thank the referee for useful suggestions
for major improvement of the original manuscript.
We also thank  H. Ito for helpful discussions.
This work is supported in part by Ministry of Education, 
Culture, Sports,Science, and Technology (MEXT) 
Research Activity Start-up 2284007 (NK).



%




\end{document}